\documentclass[aps,prb,twocolumn,groupedaddress,showpacs]{revtex4}
\usepackage{graphicx}
\usepackage{amssymb}
\usepackage{bm}
\usepackage{subfigure}
\usepackage{graphpap}
\begin{document}
\title{VO$_2$ as a natural optical metamaterial}
\author{Miller Eaton}
\affiliation{Istituto Nanoscienze CNR-NANO Centro S3, via Campi 213A, I-41125 Modena, Italy
             and Department of Physics, Southern Illinois University, Carbondale, IL USA}
\author{Alessandra Catellani}
\affiliation{Istituto Nanoscienze CNR-NANO Centro S3, via Campi 213A, I-41125 Modena, Italy}
\author{Arrigo Calzolari}
\email[corresponding author: ]{calzolari.arrigo@nano.cnr.it}
\affiliation{Istituto Nanoscienze CNR-NANO Centro S3, via Campi 213A, I-41125 Modena, Italy}
%

\date{\today}

\begin{abstract}
VO$_2$ is a unique phase change material with strongly anisotropic electronic properties. Recently, samples have been prepared that present a co-existence of phases and thus form metal-insulator junctions of the same chemical compound.
Using first principles calculations, the optical properties of metallic and semiconducting VO$_2$ are here discussed  to design self-contained natural optical metamaterials, avoiding coupling with other dielectric media. The analysis of the optical properties complements the experiments in the description of 
the vast change in reflectance and metallicity  for both disordered and planar compounds. 
The present results also predict  the possibility to realize ordered VO$_2$ junctions operating as efficient hyperbolic metamaterials in the THz-visible range, by simply adjusting the ratio between metallic and insulating VO$_2$ content. The possibility to  excite propagating {\em volume plasmom polariton} across the metamaterial is finally discussed.
\end{abstract}


\maketitle

\section{Introduction}
\label{sec:intro}
Optical metamaterials (OMMs) \cite{Cai:1339104} are artificially nanostructured compounds that exhibit an unconventional optical response to light that is unavailable in natural materials. Although the field of metamaterials started with the search of the so-called double negative materials (or Veselago media) \cite{0034-4885-68-2-R06},  the idea of realizing artificial materials that can be engineered to support unique electromagnetic modes in the subwavelength regime opened up new conceptual frontiers for OMMs, which rapidly surpassed the concept of negative refraction. 
This has fostered the research of a new class of tunable and active metamaterials, known as hyperbolic metamaterials (HMMs) \cite{Poddubny:2013cy}. 

HMMS are highly anisotropic uniaxial materials that derive their name from the topology of the isofrequency surface $\omega$({\bf k})=constant. In uniaxial materials, the dielectric $\hat{\epsilon}$ and the permeability $\hat{\mu}$ tensors  can be written in a diagonal form, in terms of the four fundamental parameters  ($\epsilon_{\parallel}, \epsilon_{\perp}, \mu_{\parallel}, \mu_{\perp}$), where the subscripts  $\parallel$ and $\perp$ indicate components parallel and perpendicular to the anisotropy ($z$) axis.
If one out of the four fundamental parameters is negative, the isofrequency surface opens into a hyperboloid. 
For non-magnetic materials, the choice  $\epsilon_{x,y}=\epsilon_{\perp}>0$, $\epsilon_z = \epsilon_{\parallel}<0$ corresponds to a two-fold hyperboloid and the medium is called a 
{\em type-I} metamaterial; the choice $\epsilon_{x,y}=\epsilon_{\perp} < 0$, $\epsilon_z = \epsilon_{\parallel}>0$ describes a one-fold hyperboloid and the medium is called a {\em type-II} material. 
These unusual characteristics open the way to a wide range of unprecedented applications such as negative refraction \cite{Hoffman:2007jt},  sub-diffraction imaging \cite{Jacob:06}, subwavelength waveguides \cite{Ishii:2013er}, and spontaneous emission engineering \cite{2040-8986-14-6-063001}.

Hyperbolic electric dispersion is typically realized by using artificial nanostructures, such as multilayers of alternating sub-wavelength layers of metals and dielectrics or assemblies of metallic nanowires embedded in a dielectric host. The choice of the structural parameters (e.g., thickness and number of layers) as well as of the constituting materials affects the optical and plasmonic properties of the HMM as well as the frequency range of their working application. On one hand, manufacturing of artificial HMMs is still a challenge as it requires ordered growth in the sub-wavelength scale of different materials with potentially different crystalline characteristics and/or growth conditions. On the other hand, the discovery of {\em natural} HMMs \cite{Narimanov2015,Sun:2014kg} -- i.e. uniaxial single-phase materials with opposite sign of permittivity -- would offer a viable alternative to artificial metal-dielectric composites. However, the hyperbolic requirement 
$\epsilon_{\perp} \epsilon_{\parallel}<0$ is a rare condition and only few systems (such as graphite-like materials,  cuprate and ruthenate perovskites)  have been identified as natural HMMs up to now \cite{Korzeb:2015fv}.  

\begin{figure}[!t]
\begin{center}
 \includegraphics[width=7cm]{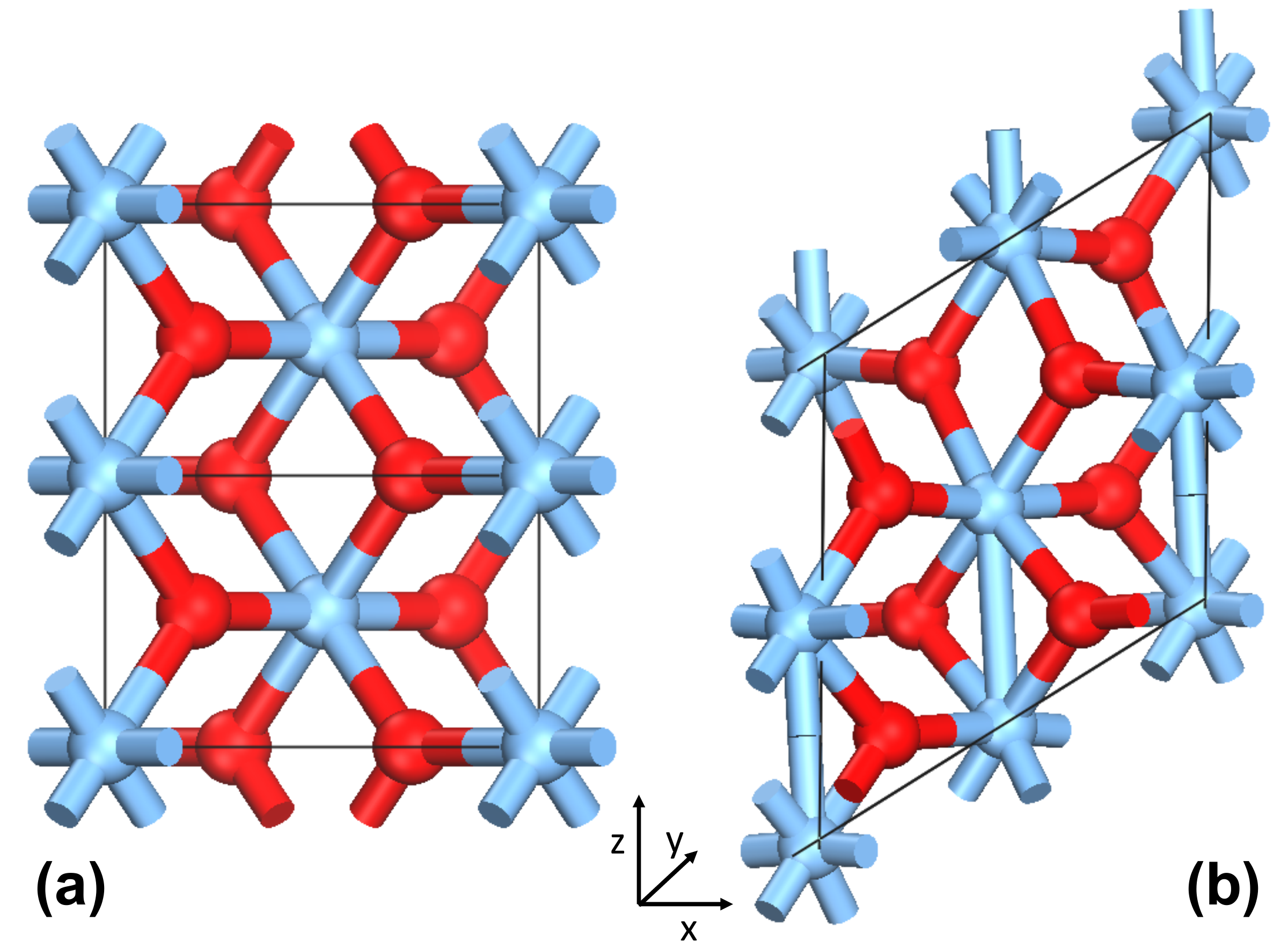}
 \caption{Crystal structure of VO$_2$ in the (a) tetragonal rutile R and (b) monoclinic M$_1$ phase. Uniaxial {\em c}-axis is aligned along z direction. The primitive cell in panel (a) is doubled along $z$ for clarity.}
 \label{fig:pos}
 \end{center}
\end{figure}

In this regard, {\em phase change materials} (PCMs) may constitute a valuable compromise for the realization of HMMs, playng with different phases, rather than different materials, thus reducing defect and strain content.
Vanadium dioxide (VO$_2$) in particular has shown promise due to its metal-to-insulator transition (MIT) \cite{RevModPhys.70.1039} at the critical temperature of ~340K \cite{Kucharczyk1979}. 
The MIT in VO$_2$ occurs when the  atoms in the metallic structure dimerize and break the crystal symmetry, resulting in a shift from a tetragonal rutile (R) phase to a low temperature monoclinic (M$_1$) phase (Fig. \ref{fig:pos}).  The optical properties differ widely between phases, making it possible to realize switchable devices based on altering the absorption, conductivity, or reflectance of the material by precise control of the system environment.  In fact, schemes are already in place to implement VO$_2$ in switching plasmonic nanoantenna, \cite{Abate2015} ultrafast light emission modulators \cite{Cueff2015}, near-field thermal transfer devices \cite{Menges2016}, and novel optical metamaterials \cite{Rensberg2016,Krishnamoorthy2014}, in connection with other dielectric oxides (e.g. sapphire \cite{Rensberg2016}, titania \cite{acsnano.6b05736}, silica\cite{Wen:2010gg}) or conducting media (e.g. Au \cite{Liu:2012fw},   Al:ZnO\cite{PhysRevApplied.8.014009}).  

Additionally, recent experiments have shown that it is possible to create VO$_2$ samples containing a co-existent mixture of both phases, 
i.e. to create a compound that combines the original metallic and dielectric behavior of the separate phases into a brand new metamaterial, with intrinsic optical properties different from the ones of the original constituents. This characteristic renders VO$_2$ a unique PCM. These VO$_2$ mixtures have often been induced through chemical doping \cite{Strelcov2016}, temperature gradients \cite{Abate2015, Savo2015}, strain \cite{Wang2016}, charge injection \cite{doi:10.1021/nl9028973}, or ion beam irradiation \cite{Rensberg2016}.  
While several experiments have been able to accurately model the optical constants of  inhomogeneous \cite{Kats2012, Qazilbash2007} or two-dimensional \cite{Rensberg2016} VO$_2$ metamaterial at a target wavelength, a comprehensive study of the effective optical properties of mixed-phase VO$_2$ across a broad range of energy and composition is still lacking.  

On the theoretical side, the majority of first principles investigations remained focused on describing the electronic band structure and characterizing the mechanism behind  MIT \cite{Brito2016, Weber2012, Eyert2011}, while modelling of their plasmonic properties uses empiric parameters extrapolated from  experiments \cite{Wen:2010gg,Yahiaoui:2017jy}.  In particular, the use of empirical data suffers from
lack of transferability, as they are strictly dependent to the specific characteristics (i.e. structural configuration, temperature, growth environment, defects, dopants, etc) of the very sample they are extracted from. Thus a fully ab initio description of the optical properties of both VO$_2$ phases at the same level of accuracy would be a dramatic improvement in the understanding of these PCM materials. First-principles simulations of VO$_2$ optical response is a dramatic challenge because of the high electron-electron correlation deriving from partially occupied and localized $V(3d)$ orbitals. This is particularly critical for the description of the monoclinc phase where the presence of the energy bandgap and of excitonic effects have to be carefully taken into account. Different approaches have been proposed so far to treat electron correlation and optical properties of VO$_2$, including dynamical mean field theory (DMFT)\cite{Brito2016} and many-boby GW Bethe-Salpheter approaches \cite{Galitski2007,Gatti2015,Continenza1999}. The inclusion of quasi-particle and local-field effects have been demonstrated to correctly reproduce the low-energy spectrum and band-gap onset of the semiconducting phase. On the other hand, such advanced many-body techniques do not give any remarkable improvement to the description of the optical properties of metallic phase, with respect to computationally simpler single-particle approaches.

Here, we provide an ab initio investigation to accurately describe the optical properties of both metallic and semi-conducting VO$_2$, and we then present a description of a combined metamaterial with a varying  degree of phase proportions and structural arrangements.  First principles calculations represent a predictive parameter-free instrument for the microscopic understanding of complex systems beyond a specific experimental condition, like applied stress, doping, defects, etc. Thus, when implementing an effective medium approximation (EMA) on the basis of the ab initio results, we are able to reproduce the vast change in reflectance and metallicity in both disordered and planar compounds, observed in the experiments 
\cite{Qazilbash2007,Qazilbash:2009gr,Rensberg2016}.   
Finally,  we investigate the hyperbolic behavior of vertically stacked  VO$_2$ herostructures and we characterize the angular behavior of so called {\em volume plasmon polaritons} \cite{Zhukovsky:13}, which may propagate across the metamaterial.

\section{Method} \label{theory}

We employ first principles DFT approaches as implemented in  the Quantum-Espresso \cite{giannozzi2009quantum} software package to calculate bulk material properties for each phase of VO$_2$.  The exchange-correlational functional is approached through a local density approximation (LDA), and norm-conserving pseudopotentials are used to describe ion cores.  Following previous theoretical investigations \cite{Brito2016, Weber2012, Eyert2011,Xiao2014}, we use experimentally obtained lattice parameters \cite{Kucharczyk1979} for both systems and optimize internal atomic positions such that atomic forces are less than 0.01 eV/\AA.  Single particle wavefunctions are expanded in a planewave  basis set up to an energy cut-off of 140 Ry, and initial calculations use uniform Monkorst-Pack grids of ($8\times8\times8$) and ($6\times6\times6$) for the respective R and M$_1$ unit cells.  We additionally employ a recent pseudohybrid Hubbard implementation of DFT+U, 
namely ACBN0 \cite{Agapito2015} that profitably corrects the DFT energy bandgap \cite{Gopal:2015bf} as well as the dielectric and vibrational properties of metal-oxides \cite{Calzolari:2013kv}. The U values for Vanadium resulting from ACBN0 cycle and used for the calculations are U(V$_d$) =0.95 eV and U(V$_d$)= 2.0 eV, for M$_1$ and R phases, respectively.
In the monoclinic phase, two sets of inequivalent oxygen are present, whose U(O$_p$)  values are 6.59 eV and 6.97 eV, while U(O$_p$)= 6.93 eV is the value for oxygen in the tetragonal phase. 

In the spirit of investigating the optical properties of mixed compounds, where the
presence of screening metallic components make correlation effects not crucial for the overall metamaterial, the optical properties are simulated at the single-particle level starting from the DFT+U description of the electronic ground state. This assures a reasonable compromise between numerical accuracy and computational cost for both M$_1$ and R phase and opens to the future possibility of treating larger simulations cells able to include dopants, defects or local disordered elements. 
In particular, the full complex dielectric function $\hat{\epsilon}={\epsilon}_r+i{\epsilon}_i$ and loss function $L=-Im[\hat{\epsilon}^{-1}(\omega)]$  are determined using the inbuilt epsilon.x code, contained within the Quantum-Espresso package, which implements  an independent-particle formulation 
of the frequency-dependent Drude-Lorentz model for the dielectric function $\hat{\epsilon}(\omega)$, where both intraband (Drude-like) and interband (Lorentz-like) contributions are explicitly considered \cite{Calzolari:2014gja}.  This approach accounts only for the electronic contribution to the dielectric function: at low energies the interaction with optical phonon modes may
affect the optical response of the system and, in a few cases, form phonon-polariton excitations not considered here. 

Once the full complex dielectric functions for  both material phases are obtained, we employ an effective medium approximation (EMA) \cite{Sihvola:1999wv} to simulate the optical constants $\tilde{\epsilon}=\tilde{\epsilon}_r+i\tilde{\epsilon}_i$ of the joint-phase metamaterial.  
Finally, the reflectance function $\widetilde{R}$ of the composite is easily obtained as: 
\begin{equation}
\widetilde{R} = \frac{(1-\tilde{n})^2+\tilde{k}^2}{(1+\tilde{n})^2+\tilde{k}^2}\\
\end{equation}
where
\begin{center}
\begin{eqnarray}
\tilde{n} &=& \{ \frac{1}{2}[(\tilde{\epsilon}_r^2+\tilde{\epsilon}_i^2)^{1/2}+ \tilde{\epsilon}_r] \} ^{1/2}\\
\tilde{k} &=& \{ \frac{1}{2}[(\tilde{\epsilon}_r^2+\tilde{\epsilon}_i^2)^{1/2}- \tilde{\epsilon}_r] \} ^{1/2}\nonumber
\end{eqnarray}
\end{center}
are the real ($\tilde{n}$) and imaginary ($\tilde{k}$) part of the effective refractive index.

\section{Results and Discussion}
\subsection{Single phase characterization}
\begin{figure}[!t]
\begin{center}
 \includegraphics[width=7cm]{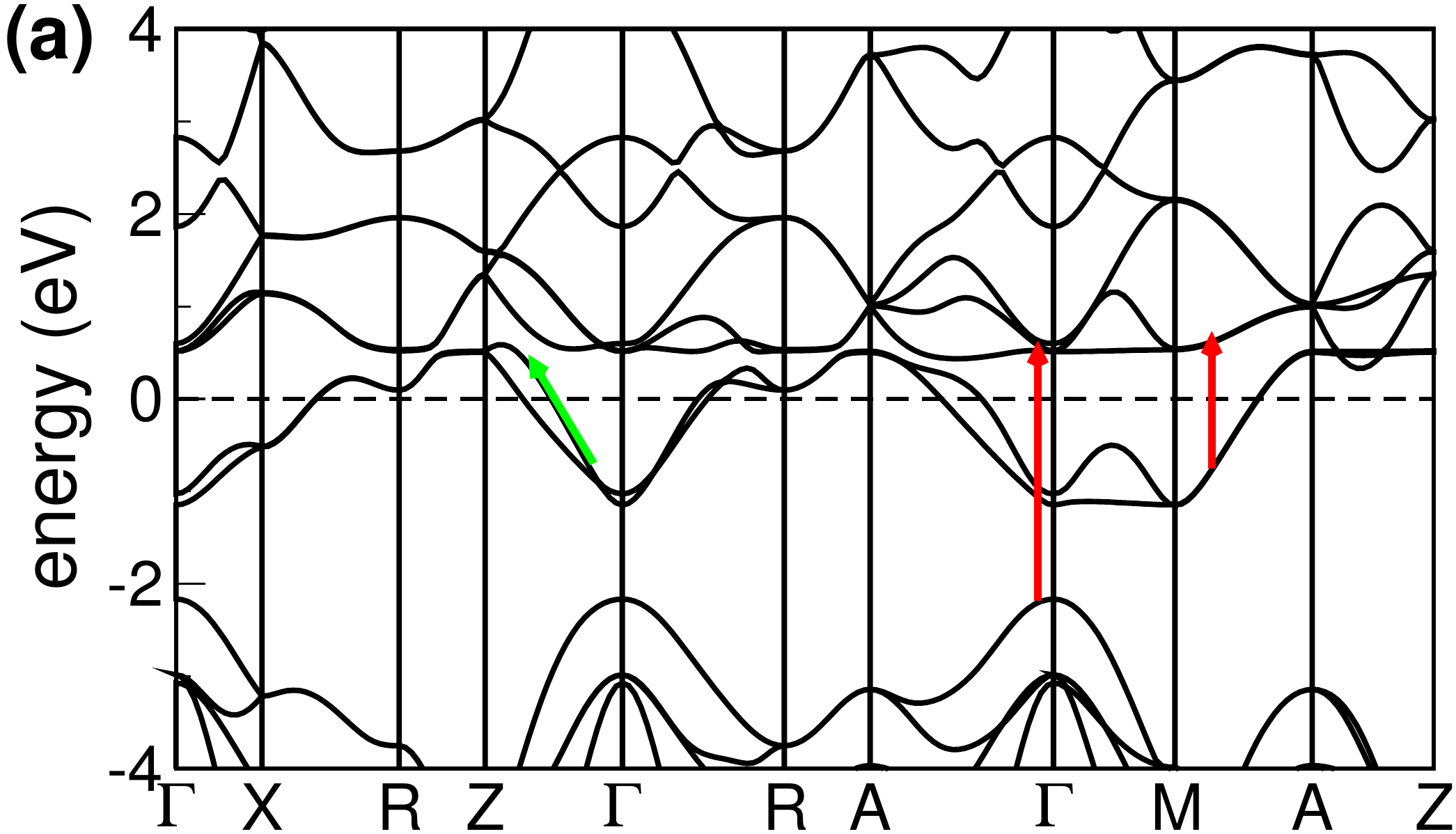}
 \includegraphics[width=7cm]{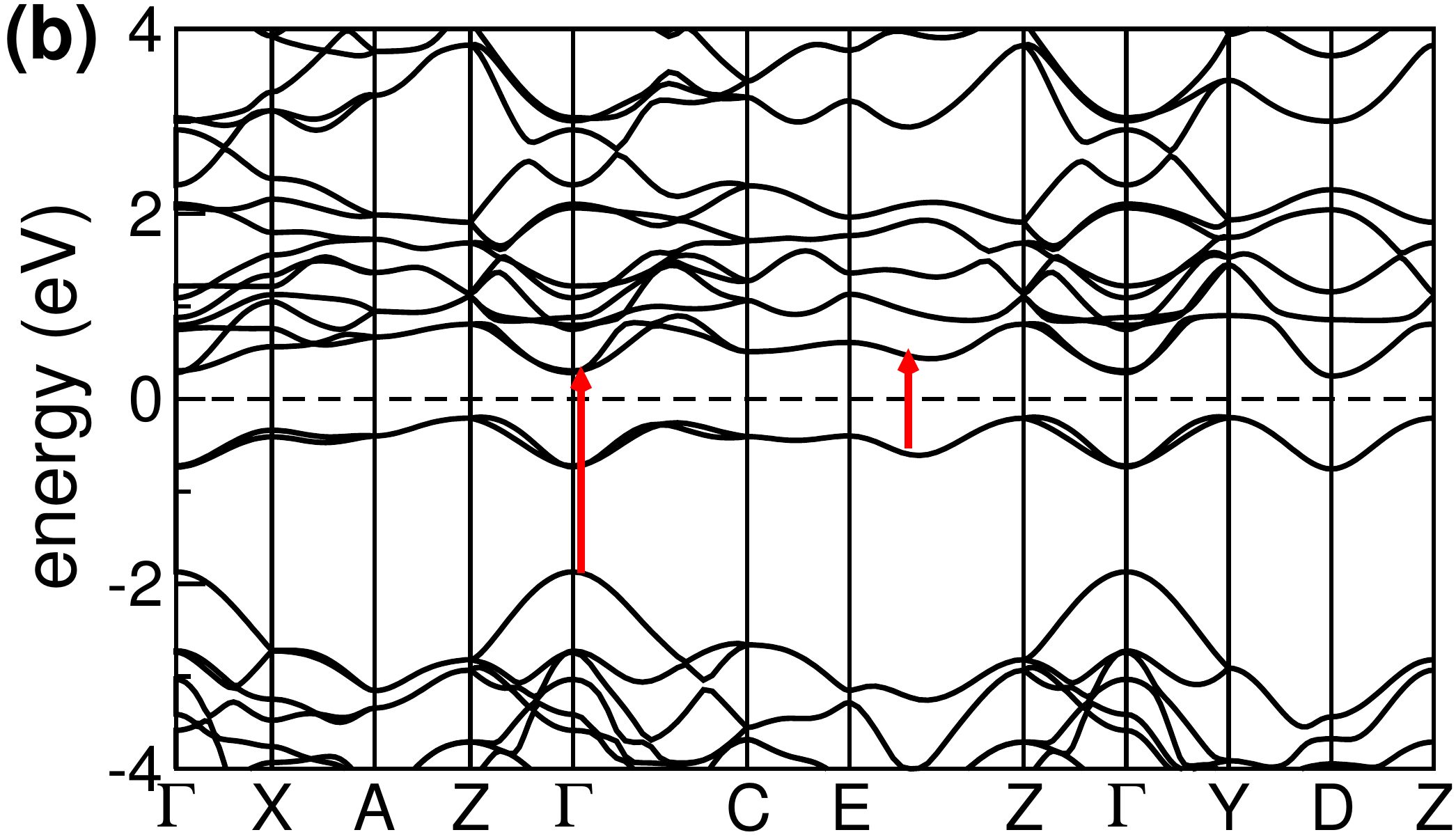}
 \caption{Electronic band structure of (a) metallic rutile  and  (b) semiconducting monoclinic VO$_2$ phase. The zero energy reference is set to the Fermi energy (dashed line) of each system. Red (green) arrows mark illustrative interband (intraband) transitions.}
 \label{fig:bands}
 \end{center}
\end{figure}

As a first step, we calculate the electronic structure of VO$_2$ for both the rutile  and monoclinic phases, whose atomic structures are shown in Fig. \ref{fig:pos} and bandstructure plots are displayed  in Fig. \ref{fig:bands}. The R phase [panel (a)] is metallic and characterized by a Fermi level (E$_F$) that overlaps 
a manyfold of bands with a predominant V$_{3d}$  character. In the octahedral coordination symmetry of rutile crystal, 
the V$_{3d}$ states are split into   $\sigma$-bonded ($e_g^{\sigma}$) orbitals, more strongly hybridized with O, and  $\pi$-bonded ($t_{2g}$) orbitals. 
The $t_{2g}$ bands can be separated into one   $a_{1g}$ band crossing the Fermi level, which derives from the overlapping of V$_{3d}$ orbitals along the rutile 
{\em c}-axis, and the remaining $e_g^{\pi}$ subbands located just above E$_F$.  
The lower-energy bands between -2 and -7 eV have a prominent O$_{2p}$ character, only partially mixed with the V states, and are separated from $t_{2g}$ bands by $\sim$1 eV. This matches the experimental values obtained by photoemission spectra \cite{Sawatzky1979,Verleur1968} and previous theoretical simulations \cite{Eyert2011,Xiao2014,Brito2016}. 

In the monoclinic phase, the crystalline distortion from the tetragonal phase induces an alternate V-V dimerization  along the {\em c}-axis [Fig. \ref{fig:pos}(b)], in agreement with the experimental findings \cite{RevModPhys.70.1039,Qazilbash2007}. The structural symmetry-breaking reflects on the bandstructure [Fig. \ref{fig:bands}(b)], which exhibits a negative energy shift of the lowest hybridized $3d$ states to just below E$_F$, resulting in a semiconductor with a narrow indirect bandgap of 0.45 eV between Z and $\Gamma$, in agreement with previous experimental \cite{Qazilbash2007,Koethe2006,Sawatzky1979} and theoretical values \cite{Brito2016, Biermann2005}.  
More specifically: due to the V-V dimerization the $a_{1g}$ band splits into bonding-antibonding subbands. This causes the lowering of the bonding-$a_{1g}$ band and the upshift of the $e_g^{\pi}$ subbands, opening the gap at the Fermi energy.
A multiplet of lower energy O$_{2p}$-derived bands appears at $\sim$-2 eV, similar to the R case.
Notably, for both phases the inclusion of the Hubbard-like correction pushes the upper V$_{3d}$ and the  O$_{2p}$ bands towards higher and lower energy, respectively, leaving the $t_{2g}$ bands close to Fermi energy almost unvaried. 
This behavior well agrees with the Mott-Peierls  character of the M$_1$ phase \cite{Qazilbash2007,PhysRevB.77.235401,Brito2016}.

\begin{figure}[!h]
\begin{center}
  \includegraphics[width=7cm]{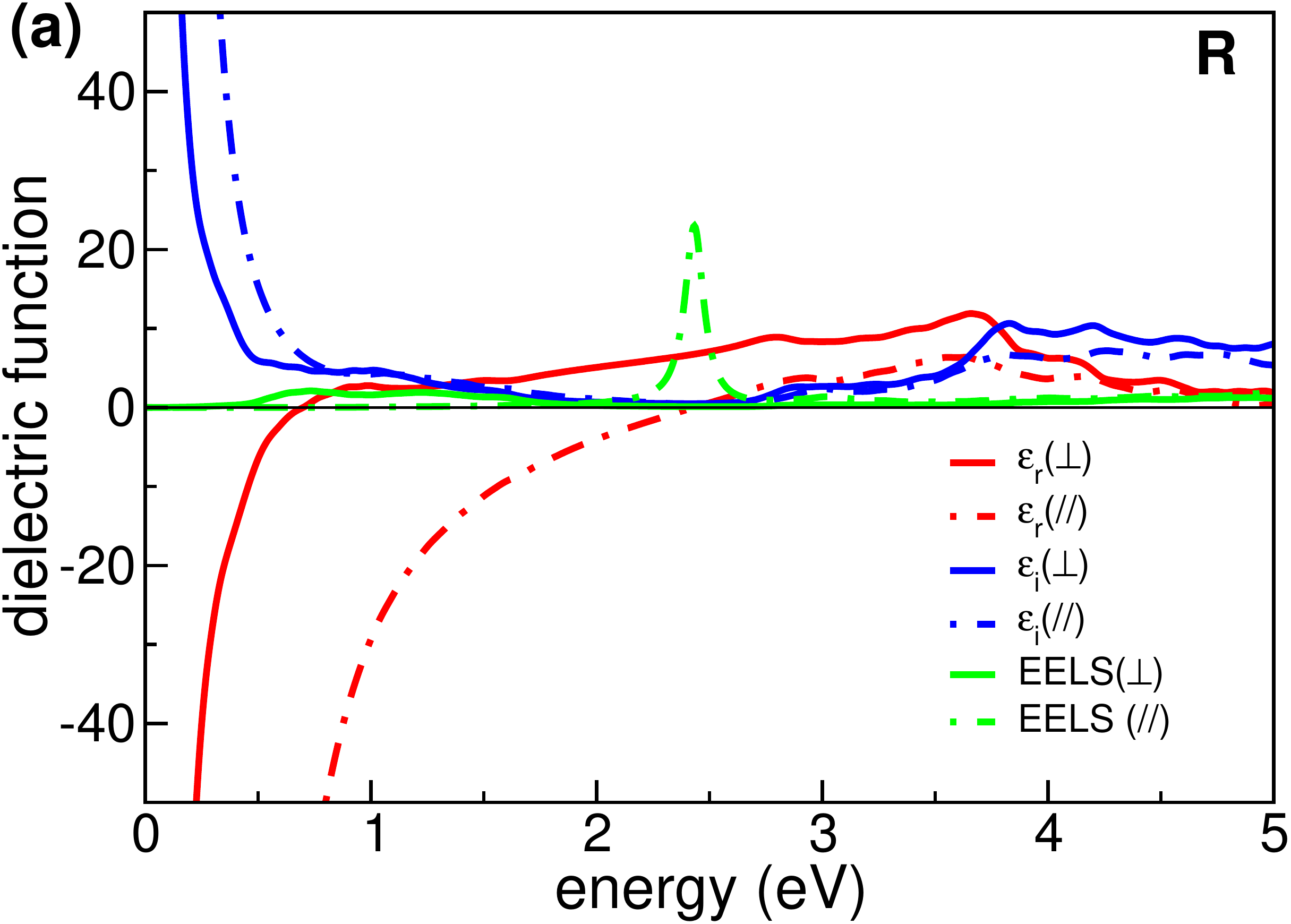}
  \includegraphics[width=7cm]{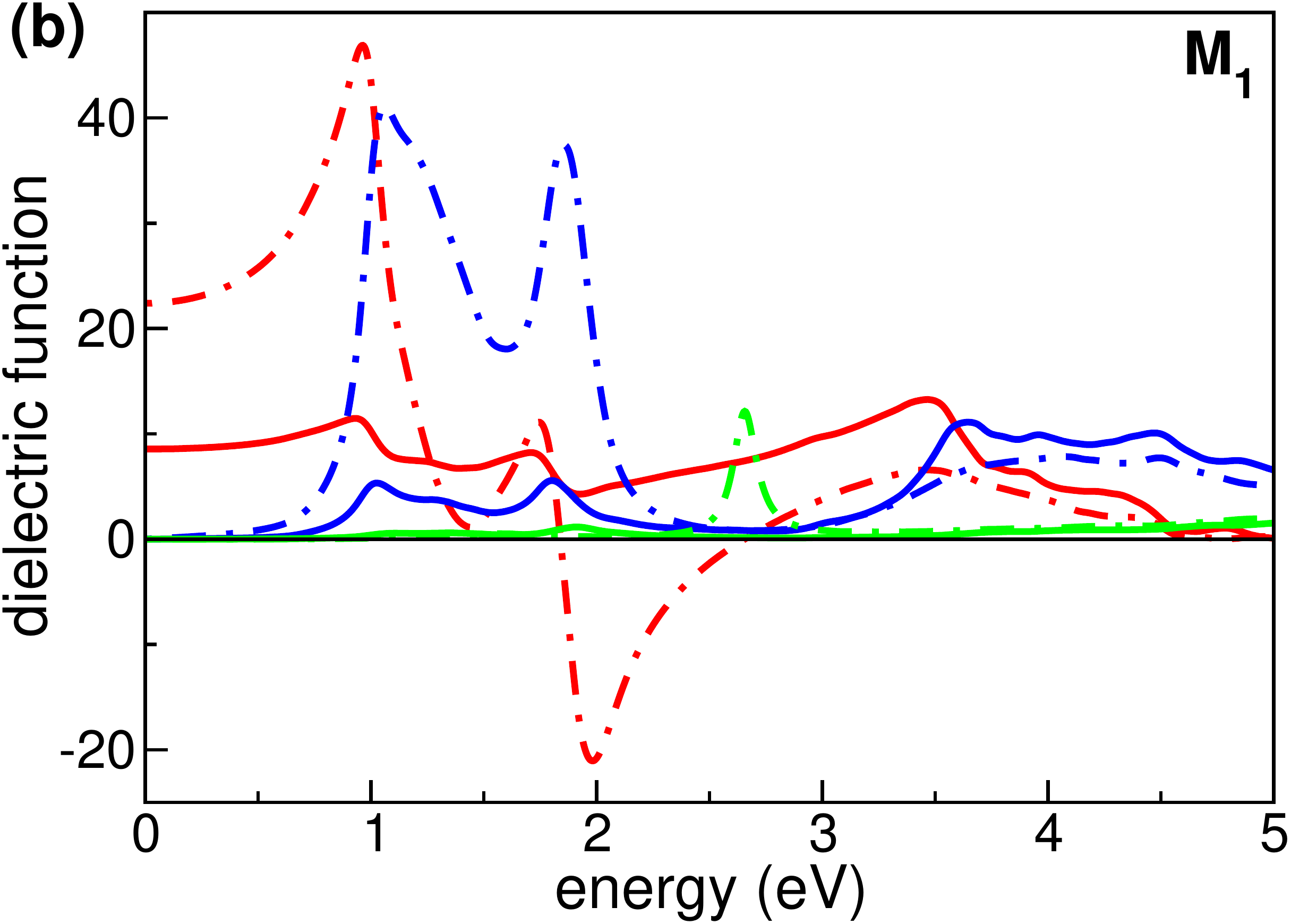}
 \caption{Real ($\epsilon_r$) and imaginary ($\epsilon_r$) part of the complex dielectric function and energy loss spectrum (EELS) for 
 VO$_2$ in (a) rutile  and (b) monoclinic  phase, calculated in the direction parallel ($//$) and perpendicular ($\perp$) to {\em c}-axis,
 as defined in Fig. \ref{fig:pos}.}
 \label{fig:epsilon}
 \end{center}
\end{figure}

VO$_2$ is highly anisotropic and this reflects on different optical response in the directions parallel or perpendicular to the {\em c}-axis of the structure (z direction in Fig. \ref{fig:pos}). Both parallel and perpendicular components are shown in Fig. \ref{fig:epsilon}. 
The rutile system is characteristic of a Drude-like picture: the divergence at zero energy and the rapid decay of the imaginary part of the dielectric function are representative of the energy damping associated with the {\em dc} electrical resistivity of  metals.
Accordingly, $\epsilon_r$ starts strongly negative at low energy and becomes positive at the crossover energy  $E_p=2.4$ eV. 
This, along with the condition  $\epsilon_i\approx0$, results in a peak in the loss function (green curves), which can be directly compared with experiments. 
The simulated crossover energy well reproduces the experimental value
of 2.75 eV detected in EELS measurements \cite{Qazilbash2006}. The peak at $E_p=2.4$ eV is interpreted as a {\em screened plasmon} excitation, typical of transition metals, including Ag and Au \cite{rocca1995}. The origin of this plasmonic peak is associated to the coexistence of intra- and interband transitions [see, e.g., green and red arrows in Fig. \ref{fig:bands}(a)].  Free electron oscillations stem from the intraband transitions of $a_{1g}$ bands that crosses the Fermi level, giving rise to the Drude-tail of $\epsilon_r$ for E$\rightarrow0$. At higher excitation energies, the activation of interband transitions between, e.g., occupied $O(2p)$ and empty $V(t_{2g})$ states gives a positive contribution to the (negative) Drude component of $\epsilon_r$ that becomes  positive at the crossover energy $E_p$.  This plasmon is usually labeled as {\em screened} as it involves the collective oscillation only of a fraction of the total valence electrons that can be considered as free, the rest being effectively screened by interband transitions. The broad and structured band in $\epsilon_i$ spectrum for E$>$ 2.5 eV is associated with the optical absorption of the material in the visible range and  
is due to interband transitions between occupied O$_{2p}$ and unoccupied V$_{3d}$ states. The crossover energy $E_p$ thus identifies two well distinguished optical characters for the material: reflective for $E<E_p$ and absorptive for $E>E_p$.

The dielectric function of monoclinic phase has the behavior typical of semiconductors
[Fig. \ref{fig:epsilon}(b)]: the real part of the dielectric function has finite value at $E=0$ eV, which corresponds to the (parallel and perpendicular) value of 
the static dielectric constant. The alternated V-V dimerization along the {\em c}-axis enhances the differences between the parallel and perpendicular components of the
dielectric function. While $\epsilon_{\perp}$ remains always positive over the entire considered energy range, $\epsilon_{\parallel}$ changes sign
in the visible range, exhibiting a typical Lorentz behavior. This generates the appearance of a crossover frequency at E=2.66 eV, which corresponds to a peak 
in the corresponding loss spectrum.
The imaginary parts are dominated by a double peak between 1-2 eV. The lowest energy component  corresponds to the onset of the insulating gap. The second major peak at 2.1 eV is the result of larger energy transitions between the $O(2p)$ states and the lowest conduction band at $\Gamma$ point.

\subsection{Joint-phase mixtures}
Although the optical anisotropy is a general condition in uniaxial systems, this is often not sufficient to qualify the intrinsic material as a natural HMM. 
Following the pioneering work by Qazilbash and coworkers \cite{Qazilbash2006,Qazilbash2007,Qazilbash:2009gr}, we considered mixed phase films whose relative semiconductor/metal ratio can be reversibly modified.  In particular, starting from room temperature semiconducting system they observed the appearance of metallic 
puddles  at the critical temperature. Puddles enlarge and coalesce as the temperature is further increased to form a complete metallic phase.
Provided the incident wavelength is large compared to the size of the metallic inhomogeneities, 
the mixed system can be described by an effective dielectric function within the EMA.

\begin{figure}[!t]
\begin{center}
 \includegraphics[width=7.5cm]{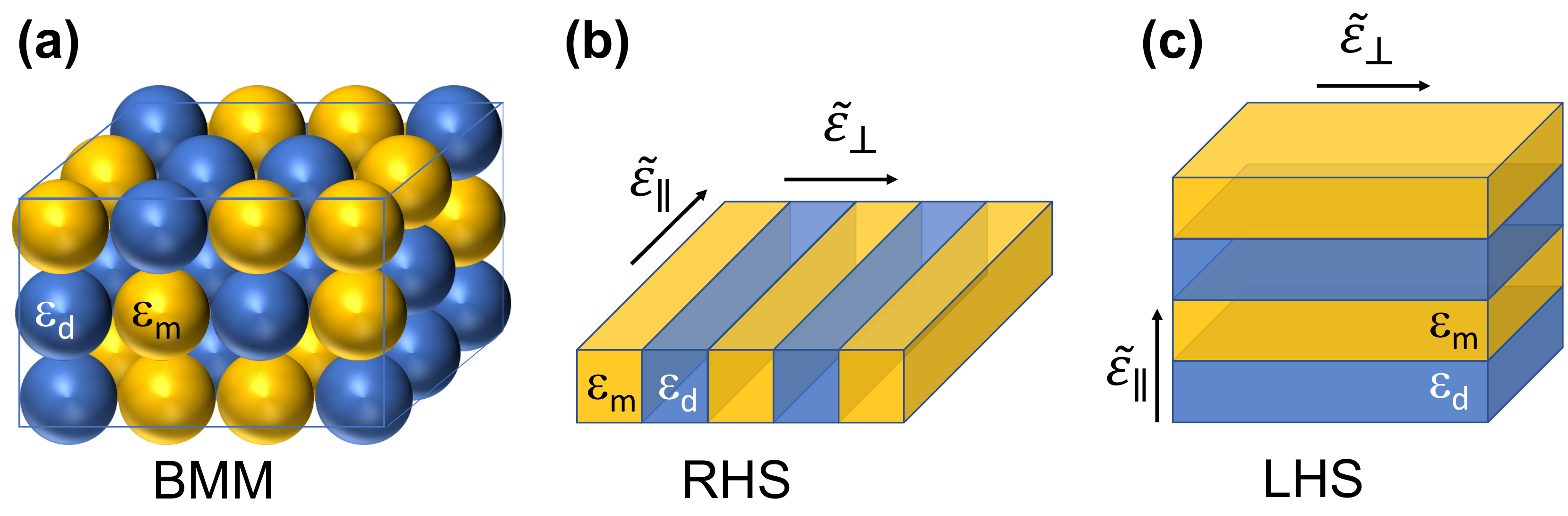}
 \caption{Scheme of the metamaterial structural models discussed in the text: (a) Bruggeman mixing model (BMM), (b) ridged heterostructure (RHS) and (c) layered heterostructures (LHS). Arrows in panels (b)-(c) identify the components
 of the dielectric function along the direction parallel ($\tilde{\epsilon}_{\parallel}$) and perpendicular ($\tilde{\epsilon}_{\perp}$) to the optical axis.}
 \label{fig:str}
 \end{center}
\end{figure}

For a homogeneous disordered systems [Fig. \ref{fig:str}(a)] we assumed a Bruggeman mixing model (BMM) \cite{Sihvola:1999wv}, where the effective dielectric function
 $\tilde{\epsilon}$ is given by solving the equation:
\begin{equation}
f\frac{\epsilon_m - \tilde{\epsilon}}{\epsilon_m + \frac{(1-q)}{q}\tilde{\epsilon}} + (1-f)\frac{\epsilon_d - \tilde{\epsilon}}{\epsilon_d + \frac{(1-q)}{q}\tilde{\epsilon}}=0 
\label{eq:bmm},
\end{equation}
where $\epsilon_m$ and $\epsilon_d$ are the complex dielectric functions of the metal rutile and dielectric monoclinic phase, respectively,
calculated from first principles \cite{note_metal}. In this case, when calculating $\tilde{\epsilon}$, the different vector components of the dielectric tensor for each individual R and M$_1$ phase are first averaged over the three spatial directions so as to simulate the random crystal orientations found in a realistic experimental environment. In Eq. (\ref{eq:bmm}) the volume fractions of the metallic and insulating phases are given by $f$ and $(1-f)$, respectively; $q$ is a depolarization factor that depends on the geometry of the phases.  If we consider an insulating film with evenly spaced, spherical metallic inclusions, the depolarization factor, $q$, is given by 1/3.  This is a valid assumption as long as the diameter of each metallic sphere does not exceed the thickness of the film.  

\begin{figure}[!t]
\begin{center}
 \includegraphics[width=7 cm]{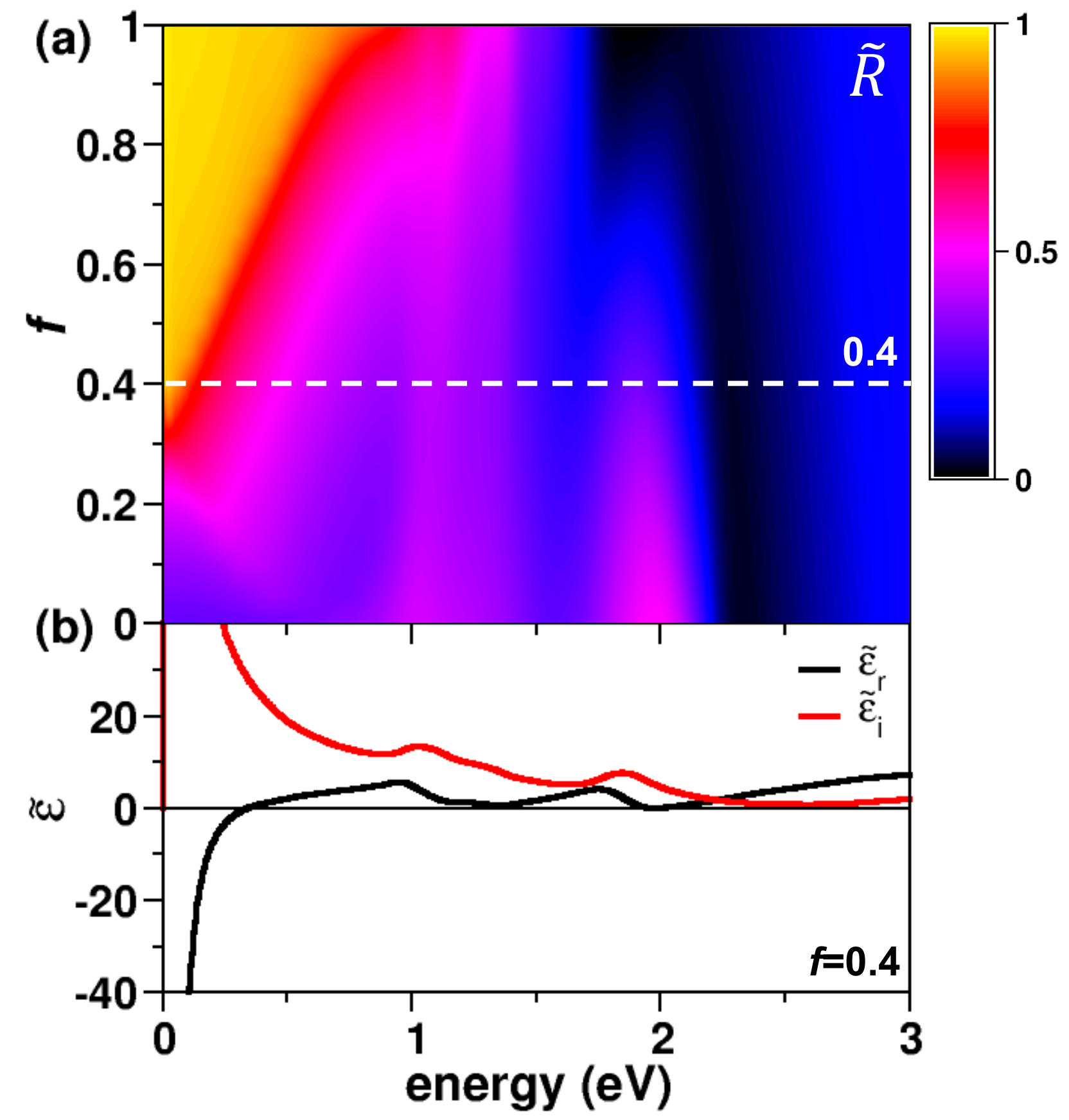}
 \caption{(a) 2D plot of reflectance $\widetilde{R}(E,f)$, as a function of the incoming radiation energy $E$ and the filling fraction $f$, according to the Bruggenman model  [Fig. \ref{fig:str}(a)].  The horizontal dashed line corresponds to the filling condition $f=0.4$, whose real ($\tilde{\epsilon}_r$) and imaginary ($\tilde{\epsilon}_i$) part of the effective dielectric function are shown in panel (b).}
 \label{fig:brug}
 \end{center}
\end{figure}

The simulated 2D plot of reflectance $\widetilde{R}(E,f)$ of the mixed phase metamaterial, as a function of the incoming radiation energy ($E$) and the filling fraction ($f$) is shown in Fig. \ref{fig:brug}(a). 
For $f=0$ and $f=1$, $\widetilde{R}$ recovers the semiconducting and metallic behavior of the M$_1$ and R phases, respectively. 
The most striking feature of the reflectance spectrum is the behavior in an energy window of 0.2 - 1.0 eV.  If we focus on a single energy value in this range, we witness the reflectance of the sample transition from  perfectly absorbing to  reflecting  as the percentage of metal is increased.  
 The modulation of $\widetilde{R}$ is in agreement with experiments that exhibit orders of magnitude shifts in reflectance of a VO$_2$ sample in the near infra-red energy regime as the material undergoes the MIT \cite{Krishnamoorthy2014,Rensberg2016,Kats2012,Qazilbash2007}. 
However, in the comparison with experiment we have to notice that our results are relative to pure VO$_2$ films.  All experiments to-date necessitate an underlying substrate that may influence the overall optical properties of the whole VO$_2$/substrate system.  
Common substrate, such as TiO$_2$ or sapphire, are themselves strongly absorbing at  singular wavelengths in THz region.  
This may explain why experiments have thus far seen near-perfect absorption at only a specific energy value and not in a larger energy range.  

For $E<1.0$ eV, while there is a point of minimal reflectance in the medium-near IR region for each filling factor, the energy value at which the film demonstrates lowest reflectance is blue-shifted as the metallic component is increased.  For $f>0.4$ the system acts as an effective metallic compound. In the energy range 1.0--2.5 eV the mixture is only partially reflecting for every semiconductor/metal composition  with  $\widetilde{R}$ being 30-50\% on average. Additionally, there is a point of even  lower reflectance for all phase proportions in the visible spectrum, at approximately 2.25 eV.  At higher energies, i.e. beyond the crossover frequency of both M$_1$ and R phases, the metamaterial is no more reflecting and the absorption processes become preponderant.
Occurrence of metallicity at  $f=0.4$  is confirmed by the plot of the real and imaginary part of the effective dielectric function, shown in Fig. \ref{fig:brug}(b) for the same $f$ value.

The spontaneous arrangement of the metallic puddles in mixed films does not allow to define preferential directions for the effective optical response. However,  throughout the spatial control of defects or strain it is possible to fabricate joint-phase metamaterials with periodic sequence of metallic and semiconducting 
sections. In particular, following the experimental work by Rensberg et {\em al} (Ref. [\citenum{Rensberg2016}]), we considered 
a ridged heterostructure (RHS) of alternating stripes of metallic and semiconducting VO$_2$ material, as shown in Fig. \ref{fig:str}b. In this case, we can easily  distinguish 
two components of the dielectric function:  the direction parallel ($\tilde{\epsilon}_{\parallel}$) or perpendicular ($\tilde{\epsilon}_{\perp}$) to the ridges.
Starting from the Bruggeman formula [Eq. (\ref{eq:bmm})] and setting the depolarization factors parallel and perpendicular to the ridges equal to $q_{\parallel}=0$ and $q_{\perp}=1$, the effective (complex)
dielectric functions are:
\begin{eqnarray}
\tilde{\epsilon}_{\parallel}(RHS)&=&f\epsilon_m+(1-f)\epsilon_d, \\
\tilde{\epsilon}_{\perp}(RHS)&=&\frac{\epsilon_m\epsilon_d}{f\epsilon_d+(1-f)\epsilon_m}\nonumber.
\label{eq:rhs}
\end{eqnarray}

\begin{figure}[!t]
\begin{center}
 \includegraphics[width=8cm]{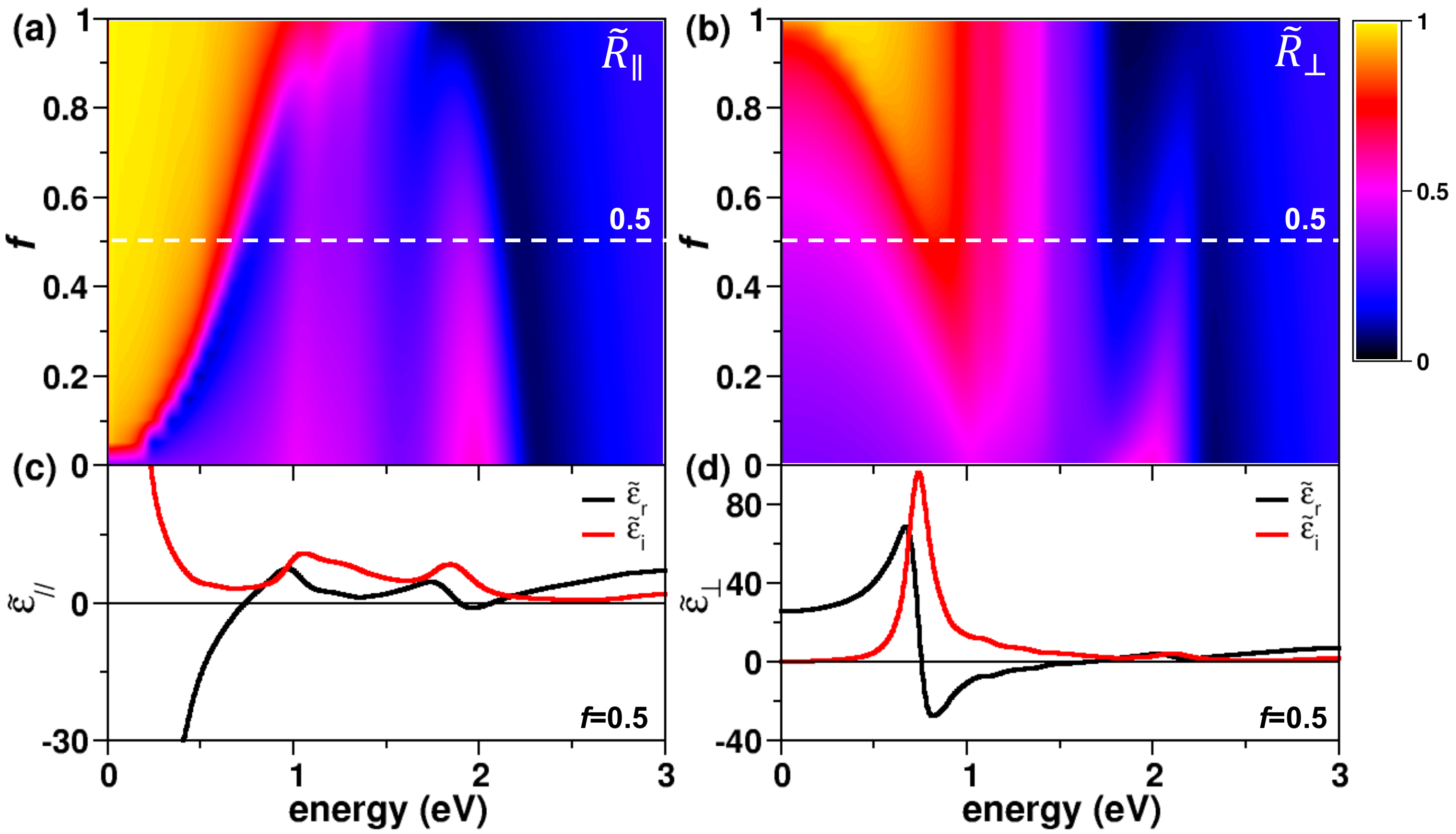}
 \caption{2D plot of reflectance $\widetilde{R}(E,f)$ as a function of the incoming radiation energy $E$ and the metallic filling fraction $f$, according to RHS model [Fig. \ref{fig:str}(b)], along the direction (a) parallel and (b) perpendicular  to VO$_2$ ridges.  Horizontal dashed lines correspond to the filling condition $f=0.5$, whose real ($\tilde{\epsilon}_r$, black line) and imaginary ($\tilde{\epsilon}_i$, red line) part of the effective dielectric function are shown in panels (c) and (d), respectively.}
 \label{fig:ridged}
 \end{center}
\end{figure}

The structural spatial anisotropy of ridged heterostructure clearly reflects on the optical response, as shown in Fig. \ref{fig:ridged}.
At low energies (E$<$ 0.5 eV), when polarized along the ridges the incoming electric field is perfectly reflected for a range of metal content.  
Conversely, in the IR-vis range the system is only partially reflecting having $\widetilde{R}_{\parallel} \approx 0.5$, due to interband absorption. As for BMM mixture, the minimum of reflectance 
is at $\sim$ 2.2 eV for all metallic contents. 
Completely different is the optical response in the direction perpendicular to the ridges: $\widetilde{R}_{\perp}$ is generally  $<0.5$ except for very high metal percentages ($>$80\%).
In particular, at low energies and low metallic content the material is essentially transparent, in agreement with the experimental findings \cite{Rensberg2016}.
Nevertheless, since the present theoretical description does not include the effect of the coupling with the lattice vibrations, we do not observe the Reststrahlen band and the modification of  reflectance at 11$\mu$m, as discussed in the original paper  \cite{Rensberg2016}.

The different reflectivity of RHS is made evident from the comparison of the effective dielectric functions [panels (c),(d)] calculated for $f=0.5$, 
which reproduces the experimental case of Ref. [\citenum{Rensberg2016}].  $\tilde{\epsilon}_{\parallel}$ has the typical fingerprints of a $d$-metal: it is dominated by a Drude-like tail for $E\rightarrow 0$ and by a crossover frequency of $\tilde{\epsilon}_{r}$ at E=0.75 eV, which corresponds to the excitation of a screened plasmon. Notably, with respect to the rutile  phase, the plasma frequency  is now strongly redshifted. This effect can be ascribed to insertions of the dielectric sections, which  reduce the overall free electron density.  The imaginary part, $\tilde{\epsilon}_{i}$, is characterized by an absorption band in the IR-vis region due to many interband transitions, e.g. between occupied $O(2p)$ or $V(a_{1g})$ and the $V(e_g^{\pi})$ states just above the Fermi Level, as pictorially indicated by red arrows in Fig. \ref{fig:bands}.  
Along the perpendicular direction, the dielectric function represents a semiconducting system, with $\tilde{\epsilon}_{r}$ converging to the static dielectric constant $\epsilon_0=25$ for $E\rightarrow 0$. Again, the presence of the metallic component increases the electrostatic screening to result in a higher value of $\epsilon_0$ with respect to the M$_1$ phase and in the shrinking of the absorption  band that now  is singly peaked around E=0.75 eV, i.e. the energy value corresponding to the plasma frequency of the parallel component.

The  directional optical response of this ridged structure indicates that ordered joint-phase  heterostructures  may sustain radiation with hyperbolic dispersion. In order to get deep insight on this 
point, we considered the standard geometry exploited in the  fabrication of HMMs given by the alternate stacking of metallic and dielectric layers [Fig. \ref{fig:str}(c)]. Actually, this kind of VO$_2$  structure has not been fabricated, yet. However, the rapid and continuous improvements in the spatially selective
control of metallic and semiconducting phases (e.g. checkerboard pattern, ridged and herringbone structures, stacked nanobeams) \cite{doi:10.1021/nl903765h,doi:10.1021/nl103925m,Kim2016,0957-4484-28-8-085701}  make the growth of VO$_2$ layered heterostructures (LHS) very reasonable in the short-medium term.
\begin{figure}[!t]
\begin{center}
 \includegraphics[width=7 cm]{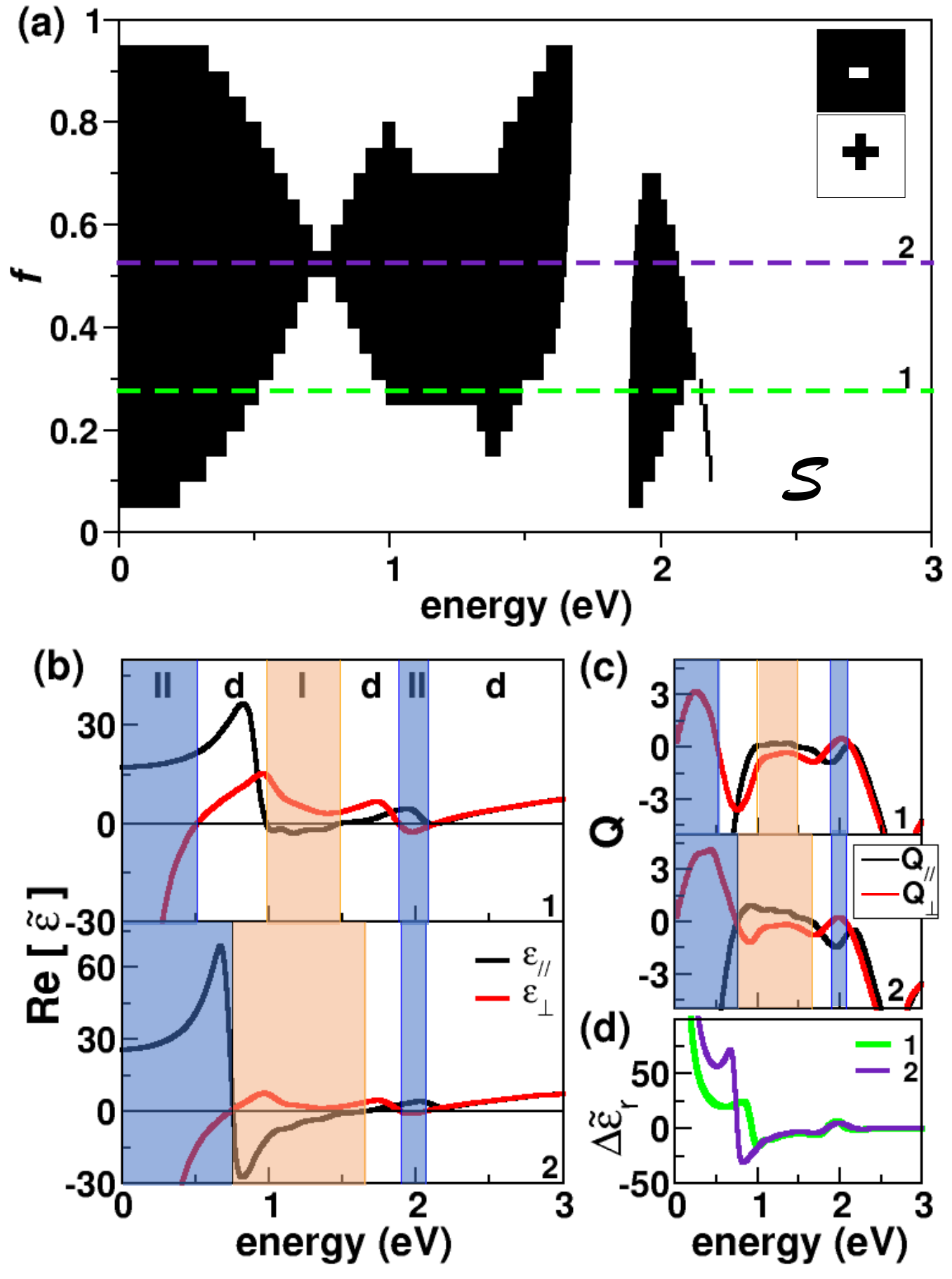}
 \caption{(a) 2D plot of the sign function $\mathcal{S}(E,f)$ as a function of the incoming radiation energy $E$ and the filling fraction $f$, resulting from the LHS model [Fig. \ref{fig:str}(c)]. Horizontal dashed lines correspond to two selected filling fractions, whose optical properties are shown in panels (b)-(d). Parallel and perpendicular components of the (b) dielectric function $\tilde{\epsilon}$ and (c) quality factor $Q$ for the two test cases  {\bf 1} and {\bf 2} selected in panel (a). (d) Strength of the dielectric anisotropy $\Delta\tilde{\epsilon}_r$.}
 \label{fig:sign}
 \end{center}
\end{figure}

Assuming the stacking direction as the anisotropy optical axis, the parallel and perpendicular components of the effective dielectric function are given by \cite{Poddubny:2013cy}:
\begin{eqnarray}
\tilde{\epsilon}_{\parallel}(LHS)&=&\frac{d_m + d_d}{d_m/\epsilon_m + d_d/\epsilon_d} \\
\tilde{\epsilon}_{\perp}(LHS)&=& \frac{\epsilon_m d_m + \epsilon_d d_d}{d_m + d_d}\nonumber,
\label{eq:lhs}
\end{eqnarray}
where  $d_m$ and $d_d$ are  thicknesses of the metallic and dielectric layers [Fig. \ref{fig:str}(c)]. The RHS and the LHS model are formally equivalent as 
$\tilde{\epsilon}_{\parallel}(RHS)=\tilde{\epsilon}_{\perp}(LHS)$ and $\tilde{\epsilon}_{\perp}(RHS)=\tilde{\epsilon}_{\parallel}(LHS)$ with $f=d_m/(d_m+d_d)$.

We analyzed the hyperbolic dispersion condition of VO$_2$ LSHs, plotting the sign function 
\begin{equation}
\mathcal{S}(E,f)=  \frac{Re[\tilde{\epsilon}_{\parallel}] Re[\tilde{\epsilon}_{\perp}]  }{ \big| Re[\tilde{\epsilon}_{\parallel}] Re[\tilde{\epsilon}_{\perp}] \big|}.
\end{equation} 
Black areas in Fig. \ref{fig:sign}(a) identify the regions where the sign function $\mathcal{S}$ is negative, i.e. where the metamaterial has  
a hyperbolic behavior: two broad interconnected regions at 0--0.7 and 0.7--1.7 eV;  one narrow region at 1.9--2.1 eV.
To investigate the characteristics of these regions we select two values of metal content, namely $f=0.25$ and $f=0.5$ (labeled {\bf 1} and {\bf 2}, respectively in Fig. \ref{fig:sign})
and we plot the corresponding  parallel and perpendicular components of the dielectric function [panel (b)]. From the comparison between the two panels it follows that each area of panel (a)
corresponds to different electrical characteristics. White sections, for which the dielectric functions have the same sign, correspond to an overall dielectric character, being both components of $\tilde{\epsilon}_r$ finite and positive. 
Black areas correspond to distinct hyperbolic types: at lower and higher energies (blue areas in Fig. \ref{fig:sign}) we see $\tilde{\epsilon}_{\parallel}>0$ and $\tilde{\epsilon}_{\perp}<0$, which is the definition of {\em type-II} HMMs, while in the intermediate region (orange areas) the condition $\tilde{\epsilon}_{\parallel}<0$ and $\tilde{\epsilon}_{\perp}>0$ identifies a {\em type-I} HMM. The energy alternation of these electrical behavior depends on the specific metal/semiconductor ratio in the sample. At low metal content ({\bf 1}) the  hyperbolic regions are separated  by dielectric energy gaps. In case {\bf 2}, the sample switches from {\em type-II} to {\em type-I} at E=0.75 eV, where both components are zero (epsilon-near-zero ENZ regime \cite{Engheta286}). 

Along the lines of Ref. [\citenum{Ishii:2013er}], we define two parameters in order to quantify the hyperbolic character of LHS, namely 
the quality factor $Q$ ,
and the strength of dielectric anisotropy  $\Delta\tilde{\epsilon}$: 
\begin{eqnarray}
Q_{j}&=&-\frac{Re[\tilde{\epsilon}_j]}{Im[\tilde{\epsilon}_j]}\\
\Delta\tilde{\epsilon}&=&Re [\tilde{\epsilon}_{\parallel}-\tilde{\epsilon}_{\perp}]\nonumber,
\label{eq:q}
\end{eqnarray}
with $j=\parallel, \perp$. The results for the key LHS systems {\bf 1} and {\bf 2} are summarized in Fig. \ref{fig:sign}(c). 
Although plotted over the entire energy range 0-3 eV for completeness, the quality factor $Q$ accounts for the energy losses only along the direction in which the material has a metallic character.  
Thus, for the perpendicular component it is meaningful in the {\em type-II}  region (blue area), while for the parallel component in the {\em type-I}  region (orange area). 
Systems with hyperbolic character  ($\mathcal{S}<0$) are considered good HMM if the maximum value of the quality factor is $Q_{max} > 3$. 
Both systems satisfy this condition at low energy ($E<0.75$ eV): they reach the maximum at $E\sim$0.3 eV, where $Q_{max}$({\bf 1})=3.1 and  $Q_{max}$({\bf 2})=4.1. 
This qualifies VO$_2$ heterostructures as good {\em type-II} HMM materials in the THz and mid-IR regions. At higher energies, 
despite a formal hyperbolic character of the system, the metal energy loss prevents any realistic application of these LHSs as good metamaterials.  

The same conclusion comes from the analysis of the strength of dielectric anisotropy [Fig. \ref{fig:sign}(d)]. 
Since the hyperbolic behavior derives from the spatial anisotropy of the optical response, the quality of HMM derives 
not only from the magnitudes of the constituent dielectric functions, but from their differences or ratios. 
As the metal ratio in the heterostructure increases, the maximum value of $\Delta\tilde{\epsilon}$ increases and diverges for $E\rightarrow0$ while its energy position is monotonically redshifted. 
This corresponds to the fact that for $E\rightarrow0$, the real part of the dielectric function in the metal component becomes so negative that there is an impedance mismatch with the other medium.
Systems with $\Delta\tilde{\epsilon}_{max} >$20--30 are considered good 
HMMs \cite{Korzeb:2015fv}. 
The complete analysis of our results show  that in the case of layered VO$_2$, this condition is reached for $f\ge0.25$, with $\Delta\tilde{\epsilon}_{max}$ 
ranging from mid-IR to THz regions. The strength of dielectric anisotropy also affects the dispersion relation of the  radiation that may propagate along the medium, as $\epsilon_{\parallel}$ and $\epsilon_{\perp}$ define the geometric characteristics of the hyperbolic isosurface equation $\frac{k_{\perp}^2}{\epsilon_{\parallel}}+ \frac{k_{\parallel}^2}{\epsilon_{\perp}}=\omega^2/c^2$.

\begin{figure}[!t]
\begin{center}
 \includegraphics[width=7.5 cm]{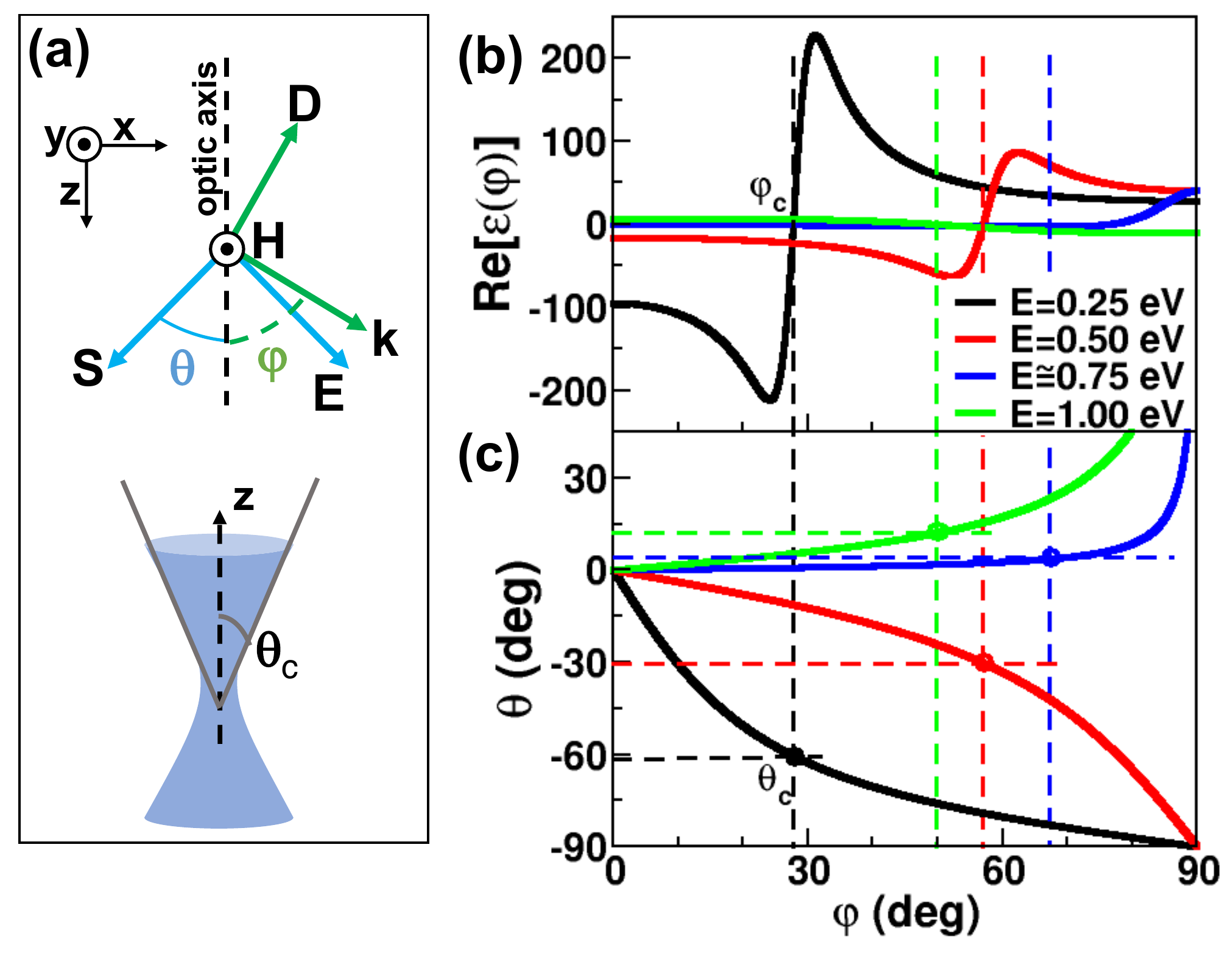}
 \caption{(a) Vector diagram for TM propagating waves and  hyperbolic dispersion isosurface corresponding to a {\em type-II} HMM.
Angular dependence of (b) the real  part of the effective function $\epsilon(\varphi)$ and (c) the $\theta$ angle between the extraordinary wave and the optical axis,
at different incoming electric field for $f=0.5$.  For each incoming energy, the vertical (horizontal) dotted lines identify the critical angle $\varphi_c$ ($\theta_c$).}
 \label{fig:ang}
 \end{center}
\end{figure}

Let us consider the radiation pattern in the case of VO$_2$ LHSs:
In general, electromagnetic waves cannot propagate in media with negative permittivity (e.g. metals). In the particular case of HMMs, the condition $\epsilon_{\parallel}\epsilon_{\perp}<0$ 
implies that the system behaves like a metal in one direction and as a dielectric in the other.  
As a consequence, waves with arbitrarily large wavevectors {\bf k} may have a propagating nature, while in isotropic materials they become evanescent due to the bound isofrequency contour (a sphere). In this sense, HMMs can be considered as plasmon-polaritonic crystals where the coupled states of light and electron density (i.e. plasmons) give rise to travelling extraordinary (TM) \cite{Belov:2003ge}  waves, known as {\em volume plasmon polaritons} (VPP) \cite{Zhukovsky:13,Ishii:2013er}.
To characterize these VPPs, we consider the electromagnetic field condition corresponding to the vector diagram shown in  Fig. \ref{fig:ang}(a), where {\bf E} is the electric-field vector, {\bf D} is the electric-displacement vector, and {\bf S} is the Poynting vector. {\bf D} lays in the principal plain containing both optical axis (aligned along $z$) and the wavevector.  $\varphi$ is the angle between the wavevector and the optical axis.
In anisotropic materials, the vectors {\bf E} and {\bf D} are not usually parallel. An immediate consequence of this is that the Poynting vector {\bf S}, which points in the direction of energy flow, and wavevector {\bf k}, directed along the wavefront normal, need not to be parallel. 

The spatial anisotropy of HMM can be expressed through an angular dependence of the dielectric function \cite{Ishii:2013er}: 
\begin{equation}
\frac{1}{\epsilon(\varphi)}=\frac{sin^2(\varphi)}{\epsilon_{\parallel}}+\frac{cos^2(\varphi)}{\epsilon_{\perp}}.
\label{eq:ang}
\end{equation}
The condition $Re[\epsilon(\varphi_c)]=0$ determines the angular boundary between the metallic and dielectric response of the metamaterial.  The critical angle $\varphi_c$ is the direction of the incident radiation that may excite the formation of a travelling plasmon-polariton wave in the system. 
If we consider the interface with a another dielectric medium (e.g. air), the wavevector at the interface follows Snell's law. Thus, the angle $\theta$ between the extraordinary wave (i.e. the Poynting vector) and the optical axis is $tan(\theta)= Re[\frac{\tilde{\epsilon}_{\perp}}{\tilde{\epsilon}_{\parallel}}] tan(\varphi)$ \cite{Ishii:2013er}.  At the critical angle $\varphi_c$,  $\theta_c$ reduces to 
the asymptotic expression $\theta_c=arctan\Big( \sqrt{-Re[\frac{\tilde{\epsilon}_{\perp}}{\tilde{\epsilon}_{\parallel}}}]\Big)$, and   represents the angle of propagation of the VPP along the metamaterial. This means
that the {\em plasmon-polariton}  propagates  throughout the {\em  volume} of the HMM along a {\em cone} 
with axis coincident to the optical axis and angle $\theta_c$ [Fig. \ref{fig:ang}(a)]. 
The number of VPP modes that can be excited depends on the actual number of metal/dielectric layers in the heterostructure. The maximum number of modes in an ideal structure is always  one less than the number of metallic layers. However, a real structure always has a maximum wavevector beyond which the VPP modes cannot exist. This happens when the size of the VPP wavevector becomes comparable to the size of the layers comprising the HMM and the modes no longer experience a homogeneous effective medium. The latter is also the validity breakdown of the effective medium approximation. 
The {\bf k} wavevector follows a truly hyperbolic surface (i.e. open form) only for the ideal case with zero absorption processes. In the case of real absorbing materials, the isosurface equation reduces to a closed form, whose final dispersion crucially depends on the imaginary part of the dielectric function.

The angular dependence of $\epsilon(\varphi)$ and the critical angle $\theta(\varphi)$ in the case of VO$_2$ LHSs ($f$=0.5) are shown in Fig.  \ref{fig:ang}(b-c), 
for different values of the incoming radiation energy. 
In the low energy region ($E<0.75$ eV), where the system has a {\em type-II} character (black and red lines), 
the absolute value of $\epsilon(\varphi)$ decreases, while $\varphi_c$ shifts to larger angles ($\varphi_c=28^{\circ}$ and $57^{\circ}$) as the value of $E$ is increased [panel (b)].
The dielectric functions look like a single dipolar resonance with a high value of the imaginary components (not shown). 
The corresponding negative critical angles [$\theta_c=-61^{\circ}$ and $-31^{\circ}$, panel (c)] indicate the propagation of lateral backward wave with respect to the interface without negative refraction \cite{Belov:2003ge}.  At $E=0.75$ eV the real part of $\epsilon(\varphi)$ is zero and  $\theta_c$ is indefinite. Approaching this critical value in the ENZ region, e.g. $E \cong 0.75$ eV (blue line),   
$\epsilon(\varphi)$ crosses the zero at  $\varphi_c=68^{\circ}$. The corresponding critical angle $\theta_c$ is very small, imparting an extreme conical focusing to the propagation wave. 
At $E=1.0$ eV the system has a {\em type-I} character (green line) with a small valued dielectric function, typical of dielectric-like HMMs. In this case the critical angle  $\theta_c$ is small and positive, which corresponds to a standard positive wave refraction within  a collimated propagation cone, and could be exploited for applications in hyperlenses.

\section{Conclusion}
We presented a first principles investigation of the optical properties of VO$_2$ and used  EMA to determine the properties of  mixed-phase metamaterials obtained by combining  different phases of the same  VO$_2$ compound, i.e. without the introduction of other different media, to form structured materials with original and tunable optoelectronic properties and (in certain cases)  with highly anisotropic permittivity.
Our results  complement and agree with experiments performed on both disordered mixtures and ordered ridged structures.  
Furthermore, we demonstrated the possibility to use a variable amount of metallic VO$_2$  content in device applications where it is necessary to control the optical properties at a target wavelength.  

Finally, we investigated the hyperbolic behavior and the plasmonic characteristics of stacked VO$_2$ heterostructures, a still unexplored possibility made feasible by the unique properties of this phase change compound. 
The possibility to realize joint-phase mixtures allows one to combine the optical constants of single phases to form a ready-made metamaterial with unforeseen properties. Indeed, this combines the structural tunability typical of artificial HMMs with reduced intermaterial interfaces typical of natural HMMs, and thus opens the way for a new class of metamaterials with controllable optical properties over a wide energy range (THz to visible).

\section{Acknowledgments}
We thank Thushari Jayasekera for her friendly support and advice. ME acknowledges the Southern Illinois University Study Abroad Program for facilitating international collaboration in addition to the SIU Chancellor's Scholar Program and Center for Undergraduate Research and Creative Activities (CURCA) for partial financial support.


\begin{thebibliography}{10}

\bibitem{Cai:1339104}
W.~Cai and V.~Shalaev,
\newblock {\em {Optical Metamaterials: Fundamentals and Applications}},
\newblock Springer, New York, NY, 2010.

\bibitem{0034-4885-68-2-R06}
S.~A. Ramakrishna,
\newblock Rep. Progr. Phys. {\bf 68}, 449 (2005).

\bibitem{Poddubny:2013cy}
A.~Poddubny, I.~Iorsh, P.~Belov, and Y.~Kivshar,
\newblock Nature Phot. {\bf 7}, 958 (2013).

\bibitem{Hoffman:2007jt}
A.~J. Hoffman, L.~Alekseyev, S.~S. Howard, K.~J. Franz, D.~Wasserman, V.~A.
  Podolskiy, E.~E. Narimanov, D.~L. Sivco, and C.~Gmachl,
\newblock Nature Materials {\bf 6}, 946 (2007).

\bibitem{Jacob:06}
Z.~Jacob, L.~V. Alekseyev, and E.~Narimanov,
\newblock Opt. Expr. {\bf 14}, 8247 (2006).

\bibitem{Ishii:2013er}
S.~Ishii, A.~V. Kildishev, E.~Narimanov, V.~M. Shalaev, and V.~P. Drachev,
\newblock Laser {\&} Phot. Rev. {\bf 7}, 265 (2013).

\bibitem{2040-8986-14-6-063001}
C.~L. Cortes, W.~Newman, S.~Molesky, and Z.~Jacob,
\newblock J. Optics {\bf 14}, 063001 (2012).

\bibitem{Narimanov2015}
E.~E. Narimanov and A.~V. Kildishev,
\newblock Nature Phot. {\bf 9}, 214 (2015).

\bibitem{Sun:2014kg}
J.~Sun, N.~M. Litchinitser, and J.~Zhou,
\newblock ACS Photonics {\bf 1}, 293 (2014).

\bibitem{Korzeb:2015fv}
K.~Korzeb, M.~Gajc, and D.~A. Pawlak,
\newblock Opt. Exp. {\bf 23}, 25406 (2015).

\bibitem{RevModPhys.70.1039}
M.~Imada, A.~Fujimori, and Y.~Tokura,
\newblock Rev. Mod. Phys. {\bf 70}, 1039 (1998).

\bibitem{Kucharczyk1979}
D.~Kucharczyk and T.~Niklewski,
\newblock J. Appl. Cryst. {\bf 12}, 370 (1979).

\bibitem{Abate2015}
Y.~Abate, R.~E. Marvel, J.~I. Ziegler, S.~Gamage, M.~H. Javani, M.~I. Stockman,
  and R.~F. Haglund,
\newblock Sci. Rep. {\bf 5}, 13997 (2015).

\bibitem{Cueff2015}
S.~Cueff, D.~Li, Y.~Zhou, F.~J. Wong, J.~A. Kurvits, S.~Ramanathan, and R.~Zia,
\newblock Nature Commun. {\bf 6}, 8636 (2015).

\bibitem{Menges2016}
F.~Menges, M.~Dittberner, L.~Novotny, D.~Passarello, S.~S.~P. Parkin,
  M.~Spieser, H.~Riel, and B.~Gotsmann,
\newblock Appl. Phys. Lett. {\bf 108} (2016).

\bibitem{Rensberg2016}
J.~Rensberg, S.~Zhang, Y.~Zhou, A.~S. McLeod, C.~Schwarz, M.~Goldflam, M.~Liu,
  J.~Kerbusch, R.~Nawrodt, S.~Ramanathan, D.~N. Basov, F.~Capasso, C.~Ronning,
  and M.~A. Kats,
\newblock Nano Lett. {\bf 16}, 1050 (2016).

\bibitem{Krishnamoorthy2014}
H.~N.~S. Krishnamoorthy, Y.~Zhou, S.~Ramanathan, E.~Narimanov, and V.~M. Menon,
\newblock Appl. Phys. Lett. {\bf 104}, 2012 (2014).

\bibitem{acsnano.6b05736}
Z.~Chen, X.~Wang, Y.~Qi, S.~Yang, J.~A. N.~T. Soares, B.~A. Apgar, R.~Gao,
  R.~Xu, Y.~Lee, X.~Zhang, J.~Yao, and L.~W. Martin,
\newblock ACS Nano {\bf 10}, 10237 (2016).

\bibitem{Wen:2010gg}
Q.-Y. Wen, H.-W. Zhang, Q.-H. Yang, Y.-S. Xie, K.~Chen, and Y.-L. Liu,
\newblock Appl. Phys. Lett. {\bf 97}, 021111 (2010).

\bibitem{Liu:2012fw}
M.~Liu, H.~Y. Hwang, H.~Tao, A.~C. Strikwerda, K.~Fan, G.~R. Keiser, A.~J.
  Sternbach, K.~G. West, S.~Kittiwatanakul, J.~Lu, S.~A. Wolf, F.~G. Omenetto,
  X.~Zhang, K.~A. Nelson, and R.~D. Averitt,
\newblock Nature {\bf 487}, 345 (2012).

\bibitem{PhysRevApplied.8.014009}
J.~Rensberg, Y.~Zhou, S.~Richter, C.~Wan, S.~Zhang, P.~Sch\"oppe,
  R.~Schmidt-Grund, S.~Ramanathan, F.~Capasso, M.~A. Kats, and C.~Ronning,
\newblock Phys. Rev. Appl. {\bf 8}, 014009 (2017).

\bibitem{Strelcov2016}
E.~Strelcov, A.~Ievlev, A.~Belianinov, A.~Tselev, A.~Kolmakov, and S.~V.
  Kalinin,
\newblock Sci. Rep. {\bf 6}, 29216 (2016).

\bibitem{Savo2015}
S.~Savo, Y.~Zhou, G.~Castaldi, M.~Moccia, V.~Galdi, S.~Ramanathan, and Y.~Sato,
\newblock Phys. Rev. B {\bf 91}, 1 (2015).

\bibitem{Wang2016}
Y.~Wang, J.~Zhu, W.~Yang, T.~Wen, M.~Pravica, Z.~Liu, M.~Hou, Y.~Fei, L.~Kang,
  Z.~Lin, C.~Jin, and Y.~Zhao,
\newblock Natures Commun. {\bf 7}, 12214 (2016).

\bibitem{doi:10.1021/nl9028973}
S.~Zhang, J.~Y. Chou, and L.~J. Lauhon,
\newblock Nano Lett. {\bf 9}, 4527 (2009).

\bibitem{Kats2012}
M.~A. Kats, D.~Sharma, J.~Lin, P.~Genevet, R.~Blanchard, Z.~Yang, M.~M.
  Qazilbash, D.~N. Basov, S.~Ramanathan, and F.~Capasso,
\newblock Appl. Phys. Lett. {\bf 101} (2012).

\bibitem{Qazilbash2007}
M.~M. Qazilbash, M.~Brehm, B.-G. Chae, P.-C. Ho, G.~O. Andreev, B.-J. Kim,
  S.~J. Yun, A.~V. Balatsky, M.~B. Maple, F.~Keilmann, H.-T. Kim, and D.~N.
  Basov,
\newblock Science {\bf 318}, 1750 (2007).

\bibitem{Brito2016}
W.~H. Brito, M.~C.~O. Aguiar, K.~Haule, and G.~Kotliar,
\newblock Phys. Rev. Lett. {\bf 117}, 056402 (2016).

\bibitem{Weber2012}
C.~Weber, D.~D. O'Regan, N.~D.~M. Hine, M.~C. Payne, G.~Kotliar, and P.~B.
  Littlewood,
\newblock Phys. Rev. Lett. {\bf 108}, 1 (2012).

\bibitem{Eyert2011}
V.~Eyert,
\newblock Phys. Rev. Lett. {\bf 107}, 016401 (2011).

\bibitem{Yahiaoui:2017jy}
R.~Yahiaoui and H.~H. Ouslimani,
\newblock J. Appl. Phy. {\bf 122}, 093104 (2017).

\bibitem{Galitski2007}
V.~Galitski and Y.~B. Kim,
\newblock Phys. Rev. Lett. {\bf 99}, 266403 (2007).

\bibitem{Gatti2015}
M.~Gatti, F.~Sottile, and L.~Reining,
\newblock Phys. Rev. B {\bf 91}, 195137 (2015).

\bibitem{Continenza1999}
A.~Continenza, S.~Massidda, and M.~Posternak,
\newblock Phys. Rev. B {\bf 60}, 15699 (1999).

\bibitem{Qazilbash:2009gr}
M.~M. Qazilbash, M.~Brehm, G.~O. Andreev, A.~Frenzel, P.~C. Ho, B.-G. Chae,
  B.-J. Kim, S.~J. Yun, H.-T. Kim, A.~V. Balatsky, O.~G. Shpyrko, M.~B. Maple,
  F.~Keilmann, and D.~N. Basov,
\newblock Phys. Rev. B {\bf 79}, 075107 (2009).

\bibitem{Zhukovsky:13}
S.~V. Zhukovsky, O.~Kidwai, and J.~E. Sipe,
\newblock Opt. Exp. {\bf 21}, 14982 (2013).

\bibitem{giannozzi2009quantum}
P.~Giannozzi, S.~Baroni, N.~Bonini, M.~Calandra, R.~Car, C.~Cavazzoni,
  D.~Ceresoli, G.~L. Chiarotti, M.~Cococcioni, I.~Dabo, A.~{Dal Corso}, S.~{de
  Gironcoli}, S.~Fabris, G.~Fratesi, R.~Gebauer, U.~Gerstmann, C.~Gougoussis,
  A.~Kokalj, M.~Lazzeri, L.~Martin-Samos, N.~Marzari, F.~Mauri, R.~Mazzarello,
  S.~Paolini, A.~Pasquarello, L.~Paulatto, C.~Sbraccia, S.~Scandolo,
  G.~Sclauzero, A.~P. Seitsonen, A.~Smogunov, P.~Umari, and R.~M. Wentzcovitch,
\newblock J. Phys.: Condens. Matter. {\bf 21}, 395502 (2009).

\bibitem{Xiao2014}
B.~Xiao, J.~Sun, A.~Ruzsinszky, and J.~P. Perdew,
\newblock Phys. Rev. B {\bf 90}, 085134 (2014).

\bibitem{Agapito2015}
L.~A. Agapito, S.~Curtarolo, and M.~{Buongiorno Nardelli},
\newblock Phys. Rev. X {\bf 5}, 1 (2015).

\bibitem{Gopal:2015bf}
P.~Gopal, M.~Fornari, S.~Curtarolo, L.~A. Agapito, L.~S.~I. Liyanage, and
  M.~{Buongiorno Nardelli},
\newblock Phys. Rev. B {\bf 91}, 245202 (2015).

\bibitem{Calzolari:2013kv}
A.~Calzolari and M.~{Buongiorno Nardelli},
\newblock Sci. Rep. {\bf 3}, 2999 (2013).

\bibitem{Calzolari:2014gja}
A.~Calzolari, A.~Ruini, and A.~Catellani,
\newblock ACS Phot. {\bf 1}, 703 (2014).

\bibitem{Sihvola:1999wv}
A.~H. Sihvola,
\newblock {\em Electromagnetic Mixing Formulas and Applications},
\newblock IET, London, United Kingdom, 1999.

\bibitem{Sawatzky1979}
D.~{Sawatzky, G. A., Post},
\newblock Phys. Rev. B {\bf 20} (1979).

\bibitem{Verleur1968}
H.~W. Verleur, A.~S. Barker, and C.~N. Berglund,
\newblock Phys. Rev. {\bf 172}, 788 (1968).

\bibitem{Koethe2006}
T.~C. Koethe, Z.~Hu, M.~W. Haverkort, C.~Schler-Langeheine, F.~Venturini, N.~B.
  Brookes, O.~Tjernberg, W.~Reichelt, H.~H. Hsieh, H.~J. Lin, C.~T. Chen, and
  L.~H. Tjeng,
\newblock Phys. Rev. Lett. {\bf 97}, 1 (2006).

\bibitem{Biermann2005}
S.~Biermann, A.~Poteryaev, A.~I. Lichtenstein, and A.~Georges,
\newblock Phys. Rev. Lett. {\bf 94}, 1 (2005).

\bibitem{PhysRevB.77.235401}
B.-J. Kim, Y.~W. Lee, S.~Choi, J.-W. Lim, S.~J. Yun, H.-T. Kim, T.-J. Shin, and
  H.-S. Yun,
\newblock Phys. Rev. B {\bf 77}, 235401 (2008).

\bibitem{Qazilbash2006}
M.~M. Qazilbash, K.~S. Burch, D.~Whisler, D.~Shrekenhamer, B.~G. Chae, H.~T.
  Kim, and D.~N. Basov,
\newblock Phys. Rev. B {\bf 74}, 1 (2006).

\bibitem{rocca1995}
M.~Rocca,
\newblock Surf. Sci. Rep. {\bf 12}, 1 (1995).

\bibitem{note_metal}
Due to the correlated nature of VO$_2$, at early stage after the MIT transition
  the optical behavior of metallic puddles might be different from the those of
  pure rutile ones.\cite{Qazilbash2007} Here we did not consider these
  modifications and we assumed the rutile dielectric function as the ingredient
  of the mixing phase, for any metal content.

\bibitem{doi:10.1021/nl903765h}
A.~C. Jones, S.~Berweger, J.~Wei, D.~Cobden, and M.~B. Raschke,
\newblock Nano Lett. {\bf 10}, 1574 (2010).

\bibitem{doi:10.1021/nl103925m}
S.~Zhang, I.~S. Kim, and L.~J. Lauhon,
\newblock Nano Lett. {\bf 11}, 1443 (2011).

\bibitem{Kim2016}
M.~W. Kim, S.~S. Ha, O.~Seo, D.~Y. Noh, and B.~J. Kim,
\newblock Nano Lett. {\bf 16}, 4074 (2016).

\bibitem{0957-4484-28-8-085701}
C.~McGahan, S.~Gamage, J.~Liang, B.~Cross, R.~E. Marvel, R.~F. Haglund, and
  Y.~Abate,
\newblock Nanotechnology {\bf 28}, 085701 (2017).

\bibitem{Engheta286}
N.~Engheta,
\newblock Science {\bf 340}, 286 (2013).

\bibitem{Belov:2003ge}
P.~A. Belov,
\newblock Microwave and Optical Technology Letters {\bf 37}, 259 (2003).

\end{thebibliography}

%
\end{document}